# 5*d* transition metal oxide IrO$_2$ as a material for spin current detection

Kohei Fujiwara[1,*], Yasuhiro Fukuma[1,†], Jobu Matsuno[1,‡], Hiroshi Idzuchi[2],

Yasuhiro Niimi[2], YoshiChika Otani[1,‡,2] and Hidenori Takagi[1,3]

[1]*RIKEN Advanced Science Institute, Wako, Saitama 351-0198, Japan*

[2]*Institute for Solid State Physics, University of Tokyo, Kashiwa, Chiba 277-8581, Japan*

[3]*Department of Physics, University of Tokyo, Hongo, Tokyo 113-0033, Japan*

Magnetic Materials Laboratory, RIKEN, 2-1 Hirosawa, Wako, Saitama 351-0198, Japan

TEL: +81-48-467-9348, FAX: +81-48-462-4649

Correspondence should be addressed to K.F. (kfujiwara@sanken.osaka-u.ac.jp) and H.T. (takagi@qmat.phys.s.u-tokyo.ac.jp).

[*]Present address: The Institute of Scientific and Industrial Research, Osaka University, Ibaraki, Osaka 567-0047, Japan

[†]Present address: Frontier Research Academy for Young Researchers, Kyushu Institute of Technology, Iizuka, Fukuoka 820-8502, Japan

[‡]Present address: RIKEN Center for Emergent Matter Science (CEMS), Wako, Saitama 351-0198, Japan

**Devices based on a pure spin current (a flow of spin angular momentum) have been attracting increasing attention as key ingredients for low-dissipation electronics. To integrate such spintronics devices into charge-based technologies, an electric detection of spin current is essential. Inverse spin Hall effect (ISHE) converts a spin current into an electric voltage through spin-orbit coupling (SOC). Noble metals such as Pt and Pd, and also Cu-based alloys, owing to the large direct spin Hall effect (DSHE), have been regarded as potential materials for a spin-current injector. Those materials, however, are not promising as a spin-current detector based on ISHE. Their spin Hall resistivity $\rho_{SH}$, representing the performance as a detector, is not large enough mainly due to their low charge resistivity $\rho_C$. Here we demonstrate that heavy transition metal oxides (TMOs) can overcome such limitations inherent to metal-based spintronics materials. A binary 5$d$ TMO IrO$_2$, owing to its large resistivity as well as a large SOC associated with 5$d$ character of conduction electrons, was found to show a gigantic $\rho_{SH}$ ~ 38 $\mu\Omega$ cm at room temperature, one order of magnitude larger than those of noble metals and Cu-based alloys and even comparable to those of atomic layer thin film of W and Ta.**

SOC is a relativistic effect and relates the spin moment of electron to its orbital momentum via a momentum-dependent effective magnetic field. In the presence of SOC, charge current without any spin polarization can be converted into pure spin current and vice versa, known as DSHE and ISHE[1–18]. Since the first observation of DSHE in semiconductors[3,4], a variety of metals and semiconductors have been explored to realize efficient charge/spin conversion[5–20]. Heavy transition metals such as Pt[6–10,14], Au[9,12,13], and Pd[8,9,11] and also Cu-based alloys[15,16] were found to exhibit a particularly large spin Hall angle $\alpha_{SH}$ (the maximum yield of the charge/spin conversion), 0.01–0.1 at room temperature, owing to their pronounced SOC effects. These metals have an advantage for application as a spin injector. In addition to the large DSHE, their low electrical resistivity ($\rho_C$), typically $\rho_C = 10^{-7}$–$10^{-5}$ $\Omega$ cm, allows us to pass large charge current and hence to inject a large spin current without serious Joule heating.

Spin currents can be detected by using the ISHE. In contrast to the case for spin injector, the low $\rho_C$ of the heavy metals with a large $\alpha_{SH}$ gives disadvantages to the spin detection. In the ISHE, the electric voltage $\Delta V_{ISHE}$ generated by a spin current $I_S$ is in proportion to $\rho_C$[19,20]:

$$\Delta V_{ISHE} \propto \alpha_{SH}\rho_C I_S \approx \rho_{SH} I_S. \qquad (1)$$

Spin Hall resistivity given by $\rho_{SH} \approx \alpha_{SH}\rho_C$ determines the efficiency of spin-current detection. This means that a sensitive detection of spin current could be achieved in materials with both a large $\alpha_{SH}$ and a high $\rho_C$. In the typical heavy metals, their low $\rho_C$ impose strong constraints in tuning the value of $\rho_{SH}$ as it remains as low as less than 1 $\mu\Omega$ cm. One of the obvious approaches to improve the performance as spin detector could be to increase the $\rho_C$ by alloying. Alloying can in fact increase not only the $\rho_C$ but also the $\alpha_{SH}$ through an extrinsic spin Hall effect[21–23]. At room temperature where the inelastic scattering of charge carriers is dominant, however, it should be hard to anticipate a drastic increase of $\rho_C$ by alloying, e.g., more than an order of magnitude increase as compared with the pristine material. Therefore the improvement of $\rho_{SH}$ would not be achievable only by alloying.

In this work, we propose 5$d$ TMOs as alternative materials to realize both large $\alpha_{SH}$ and large $\rho_C$, which should enhance $\rho_{SH}$. The uniqueness of 5$d$ TMOs is characterized by the extremely strong SOC ~ 0.5–1 eV originating from the predominant 5$d$ character of the conduction band. The SOC in 5$d$ TMOs is in fact as strong as to reconstruct the electronic structures drastically, as first discussed in the case of $J_{eff}$=1/2 Mott state $Sr_2IrO_4$[24]. The analogous predominant SOC effect has been commonly observed in a broad range of Ir oxides including $Ba_2IrO_4$[25], $CaIrO_3$[26], and $Ir_2O_4$[27]. In those Ir oxides, in addition to the strong SOC, the localized character of $d$ orbitals gives rise to a moderately high charge resistivity $\rho_C$. Typical $\rho_C$ values of conductive 5$d$ Ir TMOs such as $IrO_2$ and $SrIrO_3$ at room temperature are of the order of $10^{-4}$–$10^{-3}$ Ω cm, at least one order of magnitude higher than those of normal $s$ metals. As the variety of materials continues to grow rapidly, 5$d$ TMOs offer a unique opportunity to explore the giant $\rho_{SH}$ for efficient spin detection.

We selected a simple binary oxide, rutile $IrO_2$[28–30], as the basis for our exploration. $IrO_2$ with $Ir^{4+}$ shares the same 5$d^5$ configuration[30] with many other Ir oxides, where the strong

SOC dominates the electronic states and the resultant electronic properties. This binary oxide has been long used as an electrode in various device applications, ranging from nonvolatile ferroelectric memories to electrochemical devices. The excellent electrode properties are due to the formation of a clean, well-defined interface with other materials owing to its high chemical and thermal stability, and superior barrier properties for oxygen diffusion. In such a clean interface, it may be possible to inject a spin current through a diffusion process; $IrO_2$ is therefore an ideal platform for testing the ISHE of $5d$ TMOs.

We fabricated cross-junction type devices consisting of an $IrO_2$ wire and a permalloy($Ni_{80}Fe_{20}$, Py)/Ag/Py lateral spin valve (LSV)[31] as shown in Figs. 1(a) and (b) (See also Methods). Spin currents are passed in the Ag strip by a nonlocal charge current injection ($I_C$) from the ferromagnetic Py electrode, and are perpendicularly injected in part into the $IrO_2$ wire through the diffusion [Fig 1(a)]. This effect, called spin absorption, enables us to conduct a quantitative analysis of the ISHE as have been demonstrated for various materials[7,8,17,18]. We examined potential of polycrystalline and amorphous $IrO_2$ wires as the spin-detector element. Their resistivities were $2.0 \times 10^{-4}$ Ω cm and $5.7 \times 10^{-4}$ Ω cm at 300 K, respectively, which are 1–2 orders of magnitude higher than those of metals that have been studied to date as a spin Hall material.

To check that the Ag layer is not oxidized by directly contacting with $IrO_2$, we first characterized the Ag/$IrO_2$ interface by means of interface resistance measurements[32]. The

current–voltage characteristics of the Ag/IrO$_2$ interface showed an ohmic behavior, typical of metal/metal interface. The slope of the data yielded a resistance–area product ($RA$) of 2.3 fΩ m$^2$ at 300 K, which is as low as that of the metallic, transparent Ag/Py interface ($RA$=1.9 fΩ m$^2$ at 300 K). An estimate of the thickness of the possibly oxidized Ag layer, using the resistivities of Ag$_2$O and AgO in the literature[33], never exceeded 0.1 nm. These observations imply that the Ag/IrO$_2$ interface in the present devices is very sharp as in metal/metal interfaces.

The spin absorption by IrO$_2$ across the interface was then confirmed by non-local spin-valve (NLSV) experiments[7,8,17,31] (see Supplementary Information for details). By analyzing the data based on the three-dimensional spin-diffusion model[18], we estimated a spin-diffusion length in polycrystalline IrO$_2$, $\lambda_{IrO_2}$ =3.8 nm at 300 K and 8.4 nm at 10 K, which are comparable to those of Pt (3–10 nm at 300 K[6–10,14]) and CuIr (5–20 nm at 10 K[17]) where relatively large spin Hall angles were observed. This indicates that spin scattering of IrO$_2$ is indeed very strong as expected from the 5$d$ conduction band[30].

By injecting spin currents through the spin absorption effect, we successfully observed ISHE in IrO$_2$. In Fig. 2, we plot the room-temperature ISHE signal $R_{ISHE}$ ($\equiv \Delta V_{ISHE}/I_C$) for polycrystalline (upper panel) and amorphous (lower panel) samples as a function of the magnetic field. Here the magnetic field was applied along the hard axis of the Py spin source [Fig. 1(a)], and $\Delta V_{ISHE}$ was induced along the long axis direction of the IrO$_2$ wire. Reflecting the magnetization process in hard axis of Py, $R_{ISHE}$ linearly increased up to ~1500 Oe and then saturated. The antisymmetric response to the applied magnetic field excludes any spurious effects such as anisotropic magnetoresistance of Py; the ISHE of $IrO_2$ is responsible for the resistance change ($2\Delta R_{ISHE}$) between positive and negative magnetic fields. The in-plane angular dependence of magnetic field effect on $R_{ISHE}$ was confirmed to be fully consistent with the ISHE origin[7].

Because of the large difference in $\rho_C$ between Ag and $IrO_2$, the charge current induced in the $IrO_2$ wire by the ISHE is partially shunted by the adjacent Ag layer. To take into account such geometric effects and to precisely determine $\rho_{SH}$, we adopted the three-dimensional spin-diffusion model (see Ref. 18 for details of the calculation). Using the ISHE signal $2\Delta R_{ISHE}$ and data obtained from the NLSV measurements, $\alpha_{SH}$ and $\rho_{SH}$ are respectively calculated to be 0.040 and 8.0 $\mu\Omega$ cm for polycrystalline $IrO_2$, and 0.080 and 37.5 $\mu\Omega$ cm for amorphous $IrO_2$ at 300 K.

Figure 3 summarizes spin Hall resistivity $\rho_{SH}$ and electrical resistivity $\rho_C$ for a variety of metals with a large spin Hall angle $\alpha_{SH}$ reported so far, together with that for $IrO_2$. It mimics the point of this work in that $\rho_{SH}$ of $IrO_2$ is distinctly large compared with typical heavy metals and their alloys of which $\rho_{SH}$ is ~ 0.16–1.2 $\mu\Omega$ cm[6–14,17,18], and is now comparable to the giant $\rho_{SH}$ that was recently discovered in ultrathin films of $\beta$-phase (A15 crystal structure) W[15] and Ta[16].

To investigate the nature of the ISHE in $IrO_2$, the temperature dependence of ISHE was measured for the polycrystalline sample. As shown in Fig. 4, $R_{ISHE}$ exhibits a strong variation with temperature. The reversal of field dependence at ~90 K implies the change of sign in ISHE. At 10 K, negative $\rho_{SH}$ of −6.2 μΩ cm was obtained from the field dependence of $R_{ISHE}$. The sign change is very likely due to the coexistence of different mechanisms responsible for the observed spin Hall effects[34]. Considering the presence of a certain amount of crystal defects such as grain boundaries, it is natural to suspect that extrinsic skew scattering contributes to the SHE in addition to the side jump like intrinsic effect arising from the band structure. If the sign of the extrinsic term is opposite to that of the intrinsic one, the total magnitude of the ISHEs may depend sharply on temperature.

The comparably large $\rho_{SH}$ in the amorphous form may point to that spin scattering in this compound occurs at the length scale of as short as a unit cell. Although the physics of spin scattering in such amorphous materials has not been understood yet, the choice of amorphous form should make it easier to fabricate spin Hall devices using such oxides.

In conclusion, 5$d$ TMOs are the most promising materials for spin detection in spintronics. We demonstrated that the combination of strong SOC and a moderately high $\rho_C$ is a key to improve the performance of spin detection. These achievements in the simple binary oxide manifest the potential of 5$d$ TMOs as a new class of spintronic materials. Strong SOC in this class of materials produces a rich variety of intriguing physical properties, including correlated topological insulator[35], Weyl semimetal[36] and Kitaev spin liquid[37]. Our discovery of the gigantic spin Hall resistivity may make 5$d$ TMOs even more fascinating.

**Methods**

**Device fabrication.** The Py/Ag/Py LSV with an IrO$_2$ middle wire was fabricated on a SiO$_2$/Si substrate. Au/Ti leads for electrical measurements were prepared on the substrate by a photolithography and an e-beam deposition. IrO$_2$ film was grown by a reactive sputtering with a 99.9% pure Ir target. A wire-structure of IrO$_2$ was formed using a posi-type resist patterned by an e-beam lithography. The typical wire width and thickness were 170 nm and 15 nm, respectively. During the deposition, the substrate was not heated, and the pressure was fixed at 0.7 Pa of Ar/O$_2$ (90:10) mixture gas. X-ray diffraction analysis indicated that the as-deposited IrO$_2$ film was amorphous. The amorphous film turned into polycrystalline after a post-annealing at 400°C in air. The LSV with a Py–Py separation of 700 nm (center-to-center) was then formed onto the IrO$_2$ wire by the shadow evaporation method[31,32]. To prevent the surface oxidation of Ag, a 2 nm-thick MgO protection layer was e-beam deposited on the surface without breaking vacuum. The width and thickness of wire were 170 nm and 100 nm for Ag, and 170 nm and 20 nm for Py, respectively. Their $\rho_C$ and spin diffusion length were 3.0 μΩ cm and 330 nm for Ag and 47 μΩ cm and 5 nm for Py respectively at 300 K. One of the two Py electrodes was made longer than the other [not shown in Fig. 1(b)] to make the contrast in the fields for magnetization reversal.

**Electrical measurements.** All electrical measurements were performed by a standard lock-in technique in a He flow cryostat. An a.c. excitation current $I_C$ of 200 μA and a frequency of 79

Hz were used for NLSV measurements. In ISHE measurements, $I_C$ of 400 μA was applied. We confirmed the linearity between the output voltage $\Delta V_{ISHE}$ and $I_C$ in the range of 100–400 μA.

**Acknowledgements**

We would like to thank K. Ohgushi for helpful discussions. This work was supported by Grants-in-Aid (No. 22340108, 23103518, 24224010) from MEXT, Japan.

**Author contributions**

K.F., Y.F. and J.M. designed the experiments. K.F., Y.F. and H.I. fabricated devices and collected the data. K.F. and Y.N. analyzed the data. K.F. wrote the manuscript with input from J.M. Y.O., and H.T. H.T. and Y.O. planned and supervised the project. All authors discussed the results.

**Figure captions**

**Figure 1. Inverse spin Hal effect measurements using a spin absorption effect in a lateral spin-valve geometry. a**, The structure of device used for ISHE measurements is shown schematically. A LSV with two Py electrodes (Py1: spin source, Py2: spin detector) bridged by an Ag spin transport layer was formed on an $IrO_2$ wire. Spin-polarized charge current, $I_C$, was injected along the arrow to accumulate pure spin currents in the Ag. The diffusion of spin current into the $IrO_2$ wire, namely, the spin absorption, gives rise to the ISHE in $IrO_2$. **b**, Scanning electron microscopy image of a typical device. The dotted lines indicate the side edges of the $IrO_2$ wire for clarity.

**Figure 2. Inverse spin Hal effect of $IrO_2$ at 300 K. a**, The ISHE signal $R_{ISHE}$ ($\equiv \Delta V_{ISHE}/I_C$) for a polycrystalline $IrO_2$ wire with a width $w$ of 170 nm was measured as a function of the applied magnetic field. **b**, The results for amorphous $IrO_2$ samples. Two devices with different $w$ values (100 and 170 nm) were measured. The 100 nm-width device showed larger $R_{ISHE}$ due to the increase of spin current density. $R_{ISHE}$ of the two amorphous devices are smaller than that of the polycrystalline sample shown in **a**, which is due to the large impedance mismatch between amorphous $IrO_2$ and Ag.

**Figure 3. Spin Hall resistivity.** Experimentally measured values of spin Hall resistivity $\rho_{SH}$

for various materials are plotted as a function of electrical resistivity $\rho_C$. Data include Pt[9,10] and Pd[11] evaluated at room temperature by spin pumping, Pt[14] and atomic layer thin films of $\beta$-W[15] and $\beta$-Ta[16] at room temperature by spin-transfer-torque ferromagnetic resonance, Au at 295 K using a planar Hall device[12], Ta, Nb, Pd, Mo, Pt[8], 12% Ir-doped Cu[17], 0.5% Bi-doped Cu[18] at 10 K, and polycrystalline and amorphous $IrO_2$ at 300 K (this work) by spin absorption.

**Figure 4. Temperature dependence of the ISHE signal in polycrystalline $IrO_2$.** $\rho_{SH}$ changes its sign around 90 K on decreasing temperature, suggestive of the coexistence of two spin-orbit scattering mechanisms, likely the presence of an extrinsic contribution in addition to the intrinsic contribution.

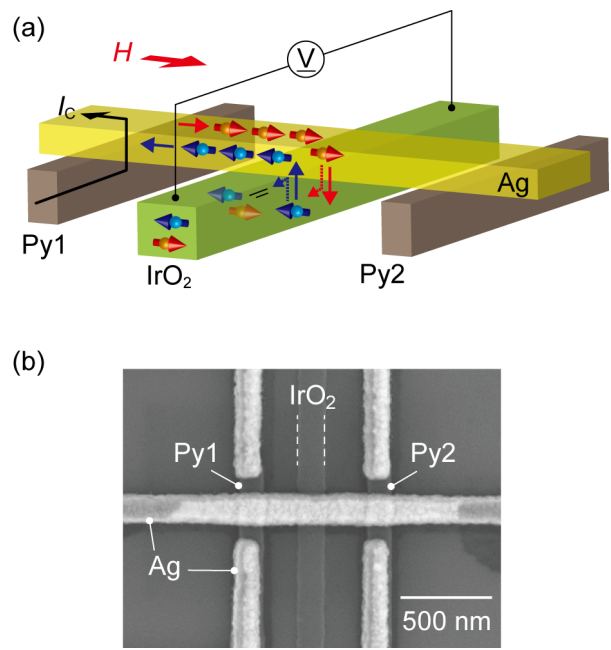

Fig. 1. K. Fujiwara *et al.*

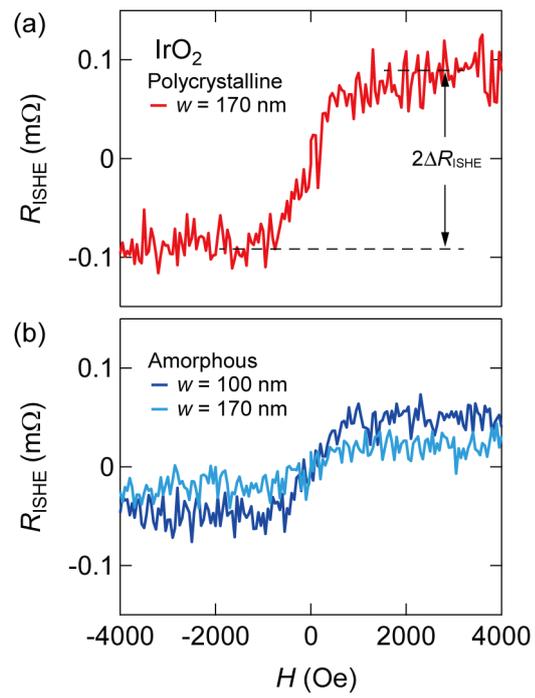

Fig. 2. K. Fujiwara *et al.*

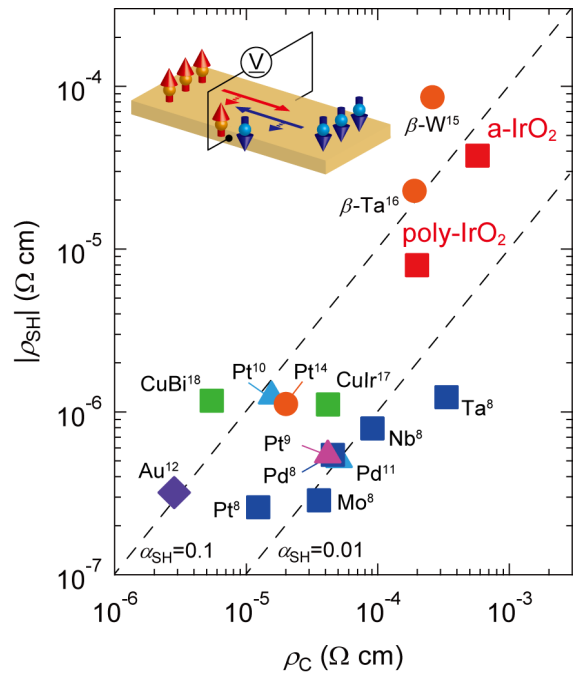

Fig. 3. K. Fujiwara *et al.*

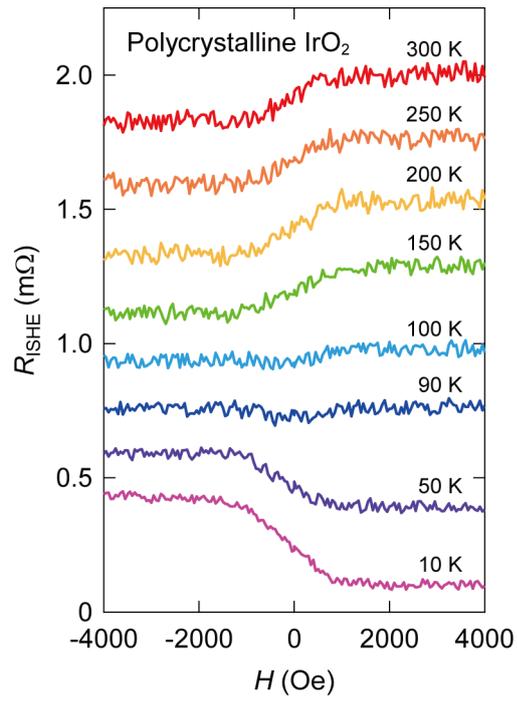

Fig. 4. K. Fujiwara *et al.*

# SUPPLEMENTARY INFORMATION

# 5*d* transition metal oxide IrO$_2$ as a material for spin current detection

**Spin absorption effect of IrO$_2$ confirmed by non-local spin-valve (NLSV) measurements**

The schematic diagram of NLSV measurements is shown in Fig. S1(a). The pure spin currents accumulated in the Ag strip by injecting spin polarized charge currents $I_C$ from Py1 to Ag is detected as a voltage $V_S$ (= $V_+ - V_-$) between Py2 ($V_+$) and Ag ($V_-$) (upper panel). The spin accumulation signal $\Delta R_S$ is given by the difference in $V_S/I_C$ ($R_S$) between parallel (high $R_S$) and antiparallel (low $R_S$) magnetizations of the two Py wires. If an IrO$_2$ middle wire is in contact with the Ag, the accumulated spin diffuses in part into it (lower panel), resulting in the reduction in $\Delta R_S$[7]. Figures S1(b) and (c) show field dependences of $R_S$ at 300 K and 10 K, respectively. The decrease of $\Delta R_S$ by the insertion of the IrO$_2$ middle wire indicates the spin absorption effect. This is more clearly demonstrated in the temperature dependence of $\Delta R_S$ with ($\Delta R_S^{\text{with}}$) and without ($\Delta R_S^{\text{ref}}$) the middle wire, displayed in Fig. S1(d). At all the temperatures, $\Delta R_S^{\text{with}}$ is systematically smaller than $\Delta R_S^{\text{ref}}$, while both are enhanced at low temperatures due to the increase of the spin-diffusion length of Ag ($\lambda_{\text{Ag}}$)[31].

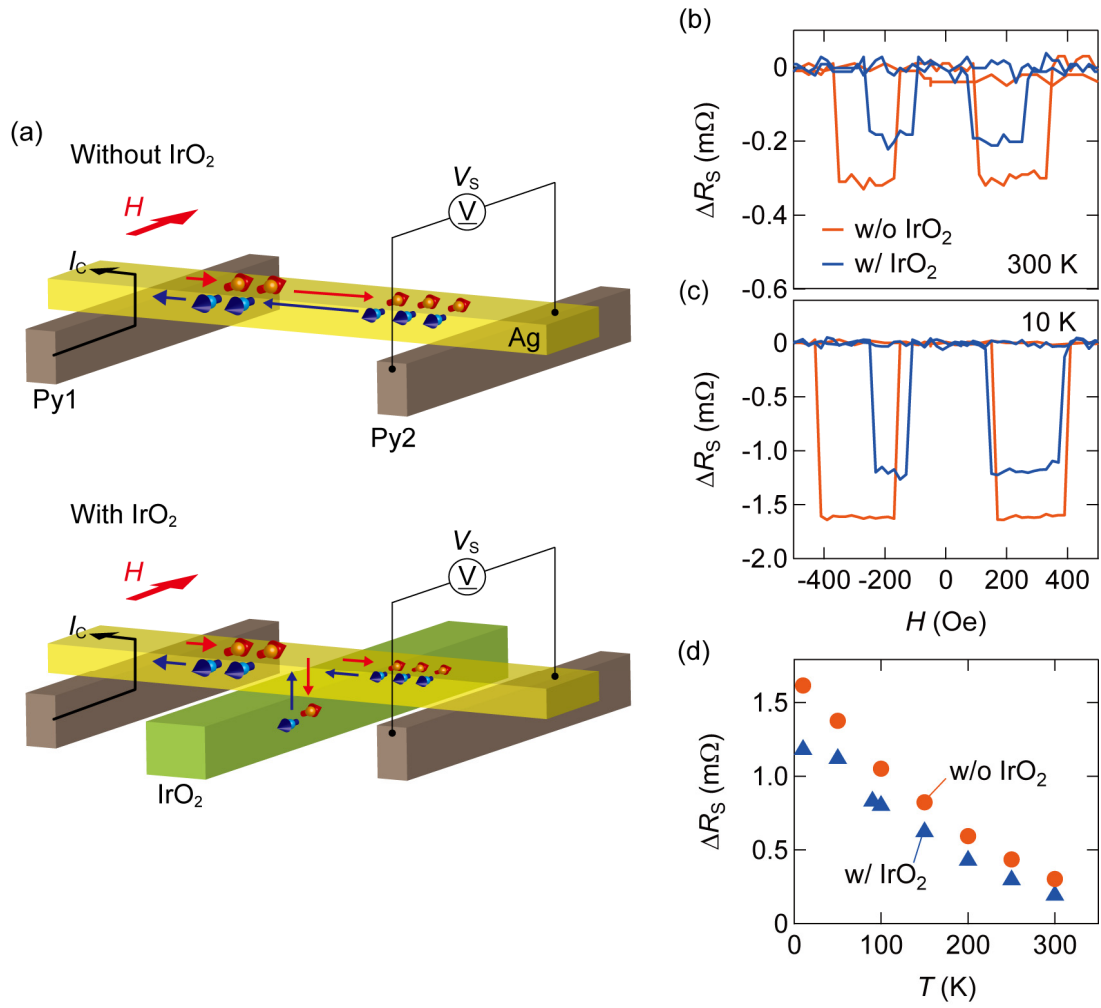

**Figure S1. Non-local spin-valve measurements. a**, The setup of measurement is schematically depicted. The magnetic field $H$ is applied along the easy axis of the Py electrodes. **b,c**, NLSV signals measured at 300 K and 10 K for Py/Ag/Py lateral spin valves without and with the polycrystalline $IrO_2$ middle wire. **d**, Temperature dependences of the NLSV signals.